\documentclass[useAMS,usenatbib,usegraphicx]{mn2e}
\usepackage{times}
\usepackage{epsfig}

\title[Globular Cluster G1 in M31]
{Structural Parameters of Mayall II = G1 in M31}
\author[Ma et al.]
{J. Ma,$^1$\thanks{E-mail:majun@vega.bac.pku.edu.cn} R. de
Grijs,$^{2,1}$ D. Chen,$^1$ S. van den Bergh,$^3$ Z. Fan,$^{1,4}$ Z.
Wu,$^1$ H. Wu,$^1$\newauthor
X. Zhou,$^1$ J. Wu, $^1$ Z. Jiang, $^1$ and J. Chen.$^1$\\
$^1$National Astronomical Observatories, Chinese Academy of
Sciences, Beijing 100012, P.R. China\\
$^2$Department of Physics \& Astronomy, The University of
Sheffield, Hicks Building, Hounsfield Road, Sheffield S3 7RH\\
$^3$Dominion Astrophysical Observatory, Herzberg Institute of
Astrophysics, National Research Council, 5071 West Saanich Road,
Victoria, BC V9E 2E7,
Canada\\
$^4$Graduate University of Chinese Academy of Sciences, A19 Yuquan
Road, Shijingshan District, Beijing 100049, P.R. China}

\date{Received; Accepted}
\pubyear{2006}
\begin{document}
\label{firstpage}
\maketitle

\begin{abstract}
Mayall II = G1 is one of the most luminous globular clusters (GCs)
known in M31. New deep, high-resolution observations with the Advanced
Camera for Surveys on the {\sl Hubble Space Telescope} are used to
provide accurate photometric data to the smallest radii yet. In
particular, we present the precise variation of ellipticity and
position angle, and of surface brightness for the core of the
object. Based on these accurate photometric data, we redetermine the
structural parameters of G1 by fitting a single-mass isotropic King
model. We derive a core radius,
$r_c=0.21\pm0.01\arcsec~(=0.78\pm0.04~\rm{pc})$, a tidal radius,
$r_t=21.8\pm1.1\arcsec~(=80.7\pm3.9~\rm{pc})$, and a concentration
index $c=\log (r_t/r_c)=2.01\pm0.02$. The central surface brightness
is 13.510 mag arcsec$^{-2}$. We also calculate the half-light radius,
at $r_h=1.73\pm0.07\arcsec(=6.5\pm0.3~\rm{pc})$.  The results show
that, within 10 core radii, a King model fits the surface brightness
distribution well. We find that this object falls in the same region
of the $M_V$ vs. $\log R_h$ diagram as $\omega$ Centauri, M54 and NGC
2419 in the Milky Way. All three of these objects have been claimed to
be the stripped cores of now defunct dwarf galaxies. We discuss in
detail whether GCs, stripped cores of dwarf spheroidals and normal
dwarf galaxies form a continuous distribution in the $M_V$ versus
$\log R_h$ plane, or if GCs and dwarf spheroidals constitute distinct
classes of objects; we present arguments in favour of this latter
view.
\end{abstract}


\section{Introduction}

Globular clusters (GCs) are effective laboratories for studying
stellar evolution and stellar dynamics. They are ancient building
blocks of galaxies, and can help us to understand the formation
and evolution of their parent galaxies. In addition, GCs exhibit
surprisingly uniform properties, suggesting a common formation
mechanism. The density distributions of most of them are well
fitted by empirical models of \citet{king62}. The closest other
populous GC system beyond the halo of our Galaxy is that of M31.
The Andromeda galaxy is the ideal nearby target for studying GCs,
since it contains more GCs than all other Local Group galaxies
combined \citep{battis87,raci91,harris91,fusi93}.

The brightest GCs in M31 are more luminous than $\omega$ Centauri,
which is the most luminous Galactic GC. Among these giants is
Mayall II = G1 (hereafter referred to as G1 for reasons of
brevity), which was first identified as a GC candidate (``Mayall
II'') by \citet{mayall53} using a Palomar 48-inch Schmidt plate
taken in 1948. It was subsequently named G1 by \citet{sarg77} in
their survey with the Kitt Peak 4-m Mayall telescope of GCs in 29
fields surrounding M31. It is located in the halo of M31, at
a projected distance of about 40 kpc from the galaxy's nucleus
(see Meylan et al. 2001). This cluster has been studied in detail
by \citet{pritchet84}, \citet{rich96}, \citet{meylan01} and
\citet{bk02}, who found that it is quite flattened, with
$\epsilon\simeq0.2$. G1 is also of interest because it may contain
a central intermediate-mass ($\sim 2 \times 10^4$ M$_\odot$) black
hole \citep{ho02,ho05}. \citet{meylan01} have pointed out that G1
is, following $\omega$ Centauri, only the second GC in which
convincing evidence for a real abundance dispersion has been seen
(although M22 and M54 are also two good candidates for a
metallicity spread; see for reviews of Sarajedini \& Layden 1995;
Da Costa \& Armandroff 1995; Monaco et al. 2004). It has
therefore been considered as the possible remnant core of a dwarf
galaxy which lost most of its envelope through tidal interactions
with M31 \citep{meylan97,meylan01}. Subsequently \citet{mackey05}
strengthened the \citet{meylan97} and \citet{meylan01} conclusion.

\citet{ho05} used the image obtained with the Advanced Camera for
Surveys (ACS) on the {\sl Hubble Space Telescope (HST)} (in fact, it
is the very same image used in this paper) to construct a radial
profile of G1, which was fitted by a non-parametric, spherical,
isotropic model to examine whether or not G1 contains a central
massive black hole. By deconvolving this better spatial resolution
image, \citet{ho05} found a bright star near the centre of G1, which
could not be detected in previous poorer resolution {\sl HST}/WFPC2
images. Therefore, the structural parameters of G1 obtained based on
this better spatial resolution image will certainly affect previous
results based on the poorer resolution {\sl HST}/WFPC2 image. This is
one of the key contributions of the present paper.

In this paper, we redetermine the structural parameters of G1
using a deep {\sl HST}/ACS image. This is at the highest
resolution yet with which this cluster has been observed; it
allows us to both probe the cluster's structure to smaller radii
than ever before and obtain the most accurate surface brightness
profile at large radii to date.

\section{Observations and Data Reductions}

\subsection{Observations and photometric data}

We searched the {\sl HST} archive and found G1 to have been observed
with the ACS/High Resolution Channel (HRC) in the F555W band
(equivalent to the Johnson $V$ filter) on 2003 October 24, as part of
programme GO-9767 (PI Gebhardt). The total integration time was
41 minutes over six exposures at three positions.  Upon retrieval from
the STScI archive, all images were processed by the standard ACS
calibration pipeline, in which bias and dark subtractions, flatfield
division, and the masking of known bad pixels are included.
Subsequently, photometric header keywords are populated. In the final
stage of the pipeline, the MultiDrizzle software is used to correct
the geometric distortion present in the HRC images. Finally, any
cosmic rays are rejected while individual images are combined into a
final single image with an exposure time of 2460s (see
Fig. \ref{fig1}). The products obtained from the STScI archive are
calibrated drizzled images, in units of counts per second. We checked
the images, and did not find saturated cluster stars.

During on-orbit operations, {\sl HST} CCD instruments are subject
to radiation damage that degrades their ability to transfer
charges. Charge transfer efficiency (CTE) degradation can lead to
photometric inaccuracy \citep[see detailed discussions
in][]{rm04,sirianni05}. However, in this paper, we did not correct for
CTE for the following reasons: (i) in the image used in this paper,
charge excesses streaming from the stellar point sources (equivalent
to the effects of bleeding, although at a much lower flux level) are
not detected, i.e., corrections for CTE are not significant; (ii) as
\citet{sirianni05} pointed out, monitoring CTE degradation is fairly
easy, but the calculation of a correction formula is more
difficult. In fact, until now, only \citet{r03} and \citet{rm04}
provided correction formulae to correct photometric losses as a
function of a source's position, flux, background, time, and aperture
size on the ACS WFC CCDs. \citet{ho05} did not correct for CTE effects
either in their analysis of the same image used in this paper.

The ACS/HRC spatial resolution is $0.025\arcsec$ pixel$^{-1}$.  This
high resolution makes two bright foreground stars appear far away from
the cluster, and hence is helpful to obtain accurate photometry of the
cluster. We used the {\sc iraf} task {\sc ellipse} to fit the image
with a series of elliptical annuli from the centre to the outskirts,
with the length of the semi-major axis increasing in steps of 8\%. The
centre coordinates of the isophotes were fixed. For our
photometry, we derived the background value as the mean of a region of
$100\times100~{\rm pixels^{2}}$ in the lower left-hand corner of the
image, the centre of which was taken 779 pixels away from the cluster
centre, and masked three areas, which were found to be disturbed
by three foreground stars. We checked the image carefully, and did not
find other obvious foreground stars. There are no obvious background
galaxies, judging from the brightnesses and extent of the objects in
the field of view. We performed the photometric calibration using the
results of \citet{sirianni05}. Magnitudes are derived in the ACS/HRC
{\sc vegamag} system. The relevant zero-point for this system is
25.255 in F555W magnitudes \citep{sirianni05}.

\subsection{Ellipticity and position angle}

\begin{figure}
\resizebox{\hsize}{!}{\rotatebox{0}{\includegraphics{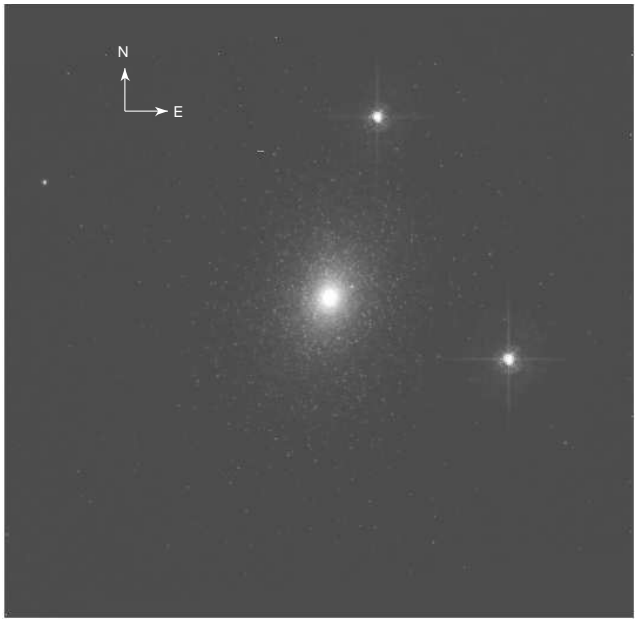}}}
\vspace{0.cm} \caption{Image of Mayall II = G1 observed with the
{\sl HST}/ACS in the F555W band. The high resolution of
$0.025\arcsec$ pixel$^{-1}$ makes two bright foreground stars
appear far away from the cluster. The image size is
$29.25\arcsec\times28.55\arcsec$.}
\label{fig1}
\end{figure}

\begin{figure}
\resizebox{\hsize}{!}{\rotatebox{0}{\includegraphics{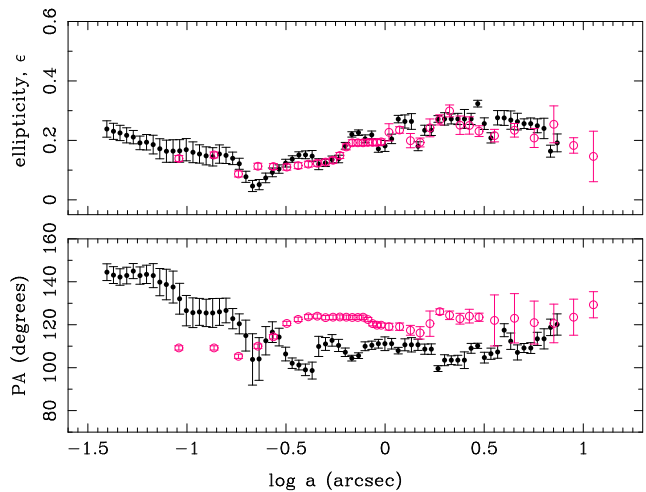}}}
\vspace{0.cm}
\caption{Ellipticity and P.A. as a function of the semimajor axis.
The filled circles are our measurements in this paper; the open
circles are from \citet{meylan01}.}
\label{fig2}
\end{figure}

Table 1 gives the ellipticity, $\epsilon=1-b/a$, and the position
angle (P.A.) as a function of the semi-major axis length, $a$,
from the centre of annulus. Position angles are measured
anti-clockwise from the vertical axis in Fig. \ref{fig1}. These
observables have also been plotted in Fig. \ref{fig2}; the
errors were generated by the {\sc iraf} task {\sc ellipse}, in
which the ellipticity errors are obtained from the internal errors
in the harmonic fit, after removal of the first and second fitted
harmonics. Beyond $a=7.385$ arcsec, the ellipticity and position
angle could not be obtained unambiguously, i.e. the fits did not
converge because of the low signal-to-noise ratio at those large
radii. The mean ellipticity is $\epsilon\simeq 0.19$, which is in
good agreement with the $\epsilon\simeq 0.2$ of \citet{meylan01}.
The ellipticity varies significantly as a function of the
cluster's semi-major axis, from a minimum of $\epsilon=0.05$ at
$a\sim0.2{\arcsec}$ to a maximum $\epsilon=0.32$ at
$a\sim2.9{\arcsec}$. This is, to within the observational
uncertainties, similar to the results of \citet{meylan01}, whose
data points are also included in Fig. \ref{fig1} for a direct
comparison (open circles). It is clear that while the general
trend of the cluster's ellipticity as a function of semimajor axis
radius is similar between the {\sl HST}/Wide Field and Planetary
Camera-2 (WFPC2)-based data of \citet{meylan01}, the improved
spatial resolution of our new {\sl HST}/ACS data allows us to
probe this trend deeper into the cluster core. Fig. \ref{fig2}
clearly shows that in the inner parts ($a<0.2{\arcsec}$) the
ellipticity increases towards smaller semimajor axis radii. This
figure also shows that uncertainties in the exact value of the
P.A. are only of secondary importance for the general trend in
ellipticity observed, given that the P.A.  determination between
\citet{meylan01} and the present paper differs by $\la 10$
degrees. There are a number of possible reasons for the offsets in
P.A. observed between these two studies, including those related
to the accuracy of the centring of our isophotes (which is linked
to the different pixel sizes), and the steps in semimajor axis
radius adopted, among others. It is of interest to note that the
high ellipticity of G1 supports the empirical rule of
\citet{bergh96} that the brightest GCs in a galaxy are also
usually the most flattened ones. The most luminous GC in M31,
037-B327, also has a high ellipticity, of $\epsilon \simeq 0.23$
\citep[see][]{ma06}, while the vast majority of M31 GCs have
ellipticities close to a median $\epsilon = 0.10$ (e.g., Lupton
1989; D'Onofrio et al. 1994; Staneva et al. 1996; Barmby et al.
2002), although some of the faintest M31 GCs show significant
flattening as well \citep{bk02}. The P.A. of the major axis is not
significantly variable for semi-major axis values $a >
\sim0.2{\arcsec}$, in agreement with \citet{meylan01}. However,
just as for the ellipticity, the P.A. also increases towards
smaller semimajor axis radii for $a <0.2{\arcsec}$.

\begin{table*}
\caption{G1: Ellipticity, $\epsilon$, and position angle (P.A.) as a
function of the semimajor axis, $a$}
\begin{tabular}{ccc|ccc}
\hline\hline
   $a$    &  $\epsilon$  &   P.A.   &   $a$    &  $\epsilon$  &   P.A. \\
 (arcsec) &              &  (deg)   & (arcsec) &      &   (deg)        \\
\hline
   0.039 & 0.238 $\pm$    0.028 & 144.5 $\pm$      3.9 &    0.583 & 0.137 $\pm$    0.013 & 110.4 $\pm$      2.8\\
   0.043 & 0.230 $\pm$    0.027 & 143.1 $\pm$      3.9 &    0.629 & 0.182 $\pm$    0.011 & 107.2 $\pm$      2.0\\
   0.046 & 0.225 $\pm$    0.026 & 142.2 $\pm$      3.8 &    0.680 & 0.220 $\pm$    0.009 & 104.4 $\pm$      1.3\\
   0.050 & 0.217 $\pm$    0.025 & 142.9 $\pm$      3.9 &    0.734 & 0.226 $\pm$    0.009 & 105.6 $\pm$      1.3\\
   0.054 & 0.211 $\pm$    0.022 & 145.0 $\pm$      3.5 &    0.793 & 0.205 $\pm$    0.013 & 110.0 $\pm$      2.1\\
   0.058 & 0.192 $\pm$    0.023 & 142.9 $\pm$      4.0 &    0.856 & 0.218 $\pm$    0.013 & 110.4 $\pm$      1.9\\
   0.063 & 0.194 $\pm$    0.026 & 143.5 $\pm$      4.3 &    0.925 & 0.171 $\pm$    0.011 & 111.2 $\pm$      2.0\\
   0.068 & 0.186 $\pm$    0.031 & 142.9 $\pm$      5.6 &    0.999 & 0.181 $\pm$    0.018 & 111.2 $\pm$      3.2\\
   0.073 & 0.172 $\pm$    0.034 & 139.8 $\pm$      6.3 &    1.078 & 0.205 $\pm$    0.016 & 111.2 $\pm$      2.5\\
   0.079 & 0.163 $\pm$    0.037 & 138.7 $\pm$      7.1 &    1.165 & 0.272 $\pm$    0.013 & 107.9 $\pm$      1.5\\
   0.085 & 0.164 $\pm$    0.038 & 137.6 $\pm$      7.4 &    1.258 & 0.264 $\pm$    0.022 & 110.7 $\pm$      2.8\\
   0.092 & 0.165 $\pm$    0.038 & 132.1 $\pm$      7.2 &    1.358 & 0.264 $\pm$    0.026 & 110.7 $\pm$      3.3\\
   0.099 & 0.169 $\pm$    0.037 & 126.5 $\pm$      7.1 &    1.467 & 0.180 $\pm$    0.015 & 110.7 $\pm$      2.6\\
   0.107 & 0.160 $\pm$    0.035 & 125.6 $\pm$      6.9 &    1.585 & 0.234 $\pm$    0.017 & 108.6 $\pm$      2.4\\
   0.116 & 0.154 $\pm$    0.031 & 126.0 $\pm$      6.3 &    1.711 & 0.234 $\pm$    0.018 & 108.6 $\pm$      2.4\\
   0.125 & 0.148 $\pm$    0.030 & 125.5 $\pm$      6.4 &    1.848 & 0.272 $\pm$    0.011 & 99.6 $\pm$      1.4\\
   0.135 & 0.145 $\pm$    0.031 & 125.5 $\pm$      6.7 &    1.996 & 0.272 $\pm$    0.021 & 103.5 $\pm$      2.6\\
   0.146 & 0.154 $\pm$    0.031 & 126.0 $\pm$      6.3 &    2.156 & 0.272 $\pm$    0.017 & 103.5 $\pm$      2.0\\
   0.157 & 0.149 $\pm$    0.027 & 126.6 $\pm$      5.7 &    2.328 & 0.272 $\pm$    0.019 & 103.5 $\pm$      2.3\\
   0.170 & 0.140 $\pm$    0.021 & 122.8 $\pm$      4.7 &    2.514 & 0.272 $\pm$    0.032 & 103.5 $\pm$      3.9\\
   0.184 & 0.121 $\pm$    0.019 & 120.4 $\pm$      4.6 &    2.716 & 0.272 $\pm$    0.019 & 109.0 $\pm$      2.3\\
   0.198 & 0.078 $\pm$    0.019 & 114.9 $\pm$      7.2 &    2.933 & 0.324 $\pm$    0.011 & 110.2 $\pm$      1.2\\
   0.214 & 0.046 $\pm$    0.019 & 103.9 $\pm$     12.0 &    3.167 & 0.256 $\pm$    0.019 & 104.7 $\pm$      2.5\\
   0.231 & 0.051 $\pm$    0.017 & 104.0 $\pm$      9.9 &    3.421 & 0.208 $\pm$    0.017 & 106.5 $\pm$      2.6\\
   0.250 & 0.074 $\pm$    0.016 & 112.5 $\pm$      6.6 &    3.695 & 0.276 $\pm$    0.024 & 107.3 $\pm$      2.9\\
   0.270 & 0.092 $\pm$    0.015 & 116.6 $\pm$      4.8 &    3.990 & 0.275 $\pm$    0.025 & 117.5 $\pm$      3.0\\
   0.291 & 0.103 $\pm$    0.014 & 114.3 $\pm$      4.1 &    4.309 & 0.268 $\pm$    0.031 & 112.3 $\pm$      3.8\\
   0.315 & 0.123 $\pm$    0.013 & 106.4 $\pm$      3.2 &    4.654 & 0.263 $\pm$    0.028 & 107.1 $\pm$      3.5\\
   0.340 & 0.137 $\pm$    0.011 & 102.1 $\pm$      2.4 &    5.026 & 0.256 $\pm$    0.016 & 109.1 $\pm$      2.1\\
   0.367 & 0.151 $\pm$    0.011 & 101.2 $\pm$      2.3 &    5.428 & 0.256 $\pm$    0.019 & 109.1 $\pm$      2.5\\
   0.397 & 0.151 $\pm$    0.013 & 99.0 $\pm$      2.7 &    5.863 & 0.249 $\pm$    0.023 & 113.4 $\pm$      3.1\\
   0.428 & 0.147 $\pm$    0.019 & 98.6 $\pm$      3.9 &    6.332 & 0.240 $\pm$    0.034 & 113.4 $\pm$      4.7\\
   0.463 & 0.120 $\pm$    0.018 & 109.9 $\pm$      4.7 &    6.838 & 0.164 $\pm$    0.020 & 118.8 $\pm$      3.8\\
   0.500 & 0.124 $\pm$    0.012 & 111.0 $\pm$      3.0 &    7.385 & 0.192 $\pm$    0.030 & 120.1 $\pm$      5.0\\
   0.539 & 0.136 $\pm$    0.012 & 112.7 $\pm$      2.7 &          &                      &                     \\
\hline
\end{tabular}
\end{table*}

\subsection{Surface brightness profile and King model fits}

After elliptical galaxies, GCs are the best understood and most
thoroughly modelled class of stellar systems. For example, a large
majority of the $\sim 150$ Galactic GCs have been fitted by the simple
models of single-mass, isotropic, lowered isothermal spheres developed
by \citet{michie63} and \citet{king66} (hereafter ``King models''),
yielding comprehensive catalogues of cluster structural parameters and
physical properties \citep[see][and references therein]{McLaughlin05}.
For extragalactic GCs, {\sl HST} imaging data have been used to fit
King models to a large number of GCs in M31 \citep[e.g.,][and
references therein]{bk02}, four GCs in M33 \citep{Larsen02}, and also
a few GCs in NGC 5128 \citep[e.g.,][and references therein]{harris02}.

\begin{figure}
\resizebox{\hsize}{!}{\rotatebox{-90}{\includegraphics{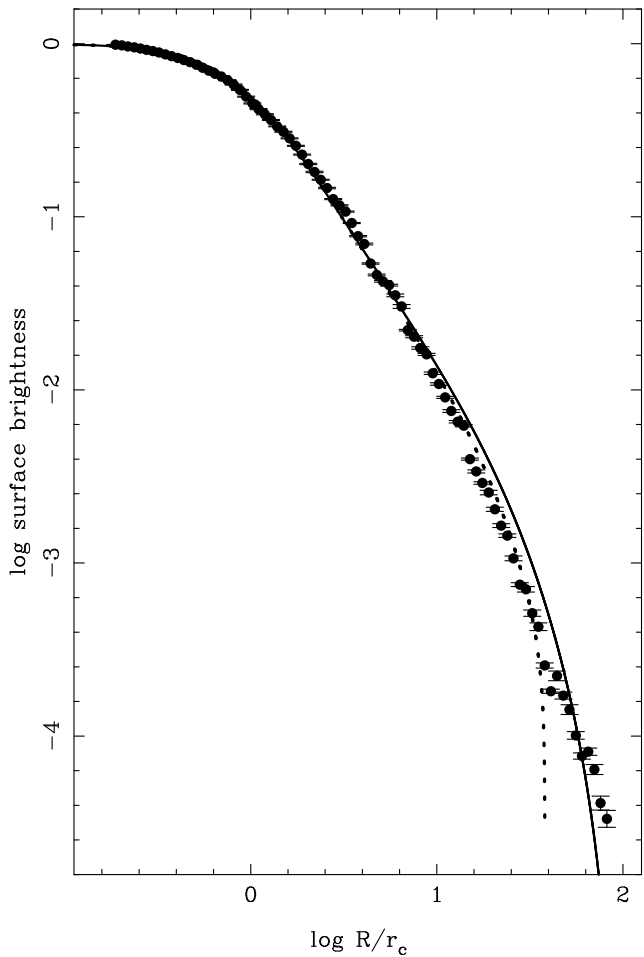}}}
\vspace{0.0cm}
\caption{Surface brightness profile of G1 measured in the F555W
passband. The solid line represents our best-fitting King model.}
\label{fig3}
\end{figure}

Table 2 lists the surface brightness, $\mu$, of G1, and its
integrated magnitude, $m$, as a function of radius. The
errors in the surface brightness were also generated by the {\sc
iraf} task {\sc ellipse}, in which they are obtained directly from
the root mean square scatter of the intensity data along the
fitted ellipse. Besides, the surface photometries at radii beyond
where the ellipticity and position angle cannot be measured, are
obtained based on the last ellipticity and position angle as the
{\sc iraf} task {\sc ellipse} is designed. The 80 points of this
observed surface brightness profile are displayed in Fig.
\ref{fig3}. We fitted King models \citep{king66} to the
surface brightness profiles. As usual, we parameterise the model
with a core radius, $r_c$, a concentration index, $c=\log
(r_t/r_c)$ (where $r_t$ is the tidal radius), and a central
surface brightness, $\mu(0)$. The fit was performed using a
nonlinear least-squares fit routine which uses the errors as
weights. We derive a core radius $r_c=0.21\pm0.01\arcsec$ and a
tidal radius $r_t=21.78\pm1.06\arcsec$, the combination of which
implies a concentration index $c=\log (r_t/r_c)=2.01\pm0.02$. The
central surface brightness is 13.510 mag arcsec$^{-2}$.  Fig.
\ref{fig3} shows the surface brightness profile and the
best-fitting King model. We also calculated the half-light radius
(the radius that contains half of the light in projection),
$r_h=1.73\pm0.07\arcsec$. Adopting a distance to M31 of 770 kpc
\citep{meylan01}, the core radius, the half-light radius and the
tidal radius are $0.78\pm0.04~\rm{pc}$, $6.5\pm0.3~\rm{pc}$ and
$80.7\pm3.9~\rm{pc}$, respectively, including their $1\sigma$
errors.

From Fig. 3, we can see that a \citet{king66} model does not fit
the observed profile of G1 very well beyond 10 core radii. We note
that we include all of the available data in our fits. Therefore, to
guarantee the proper minimum $\chi^2$ value, the observed profile at
the outer tidal region must be considered. If we do this, the fit
between 10 and 30 core radii is poor. On the other hand, the data at
radii smaller than 30 core radii can be fitted well without
considering the observed profile beyond 30 core radii, as indicated by
the dashed line in Fig. 3. We thus conclude that a single-mass
\citet{king66} model cannot fit the observed profile at the outer
regions well. G1 is only the second GC in which convincing evidence
for a real abundance dispersion has been seen, and combined with its
high brightness \citep[see details from][and references
therein]{meylan01}, it has been postulated as the possible remnant
core of a former dwarf elliptical galaxy which has lost most of its
envelope through tidal interaction with its host galaxy. It may
therefore be impossible to define its complex stellar and dynamical
properties based on simple theories for GCs, such as King models. King
models are based on the assumption that GCs are defined as
single-mass, isotropic, lowered isothermal spheres. Although this
assumption is simple, nearly all GCs can be fitted by King models
\citep[see details from][and references
therein]{bk02,Larsen02,McLaughlin05}. In fact, the structural
parameters of nearly all GCs have been and continue to be determined
on the basis of King models. However, for complicated stellar
populations such as the stripped cores of a former dwarf galaxy, King
models may not fit their profiles well in the tidal regions, due to
the stronger tidal force of the host galaxy. However, we emphasize
that \citet{meylan01} fitted the surface brightness profile of G1 with
multi-mass King models, as defined by \citet{gg79}; the result was
extremely good. \citet{meylan01} use four free parameters, in addition
to an initial mass function (IMF) exponent: (i) the core radius, (ii)
the scale velocity, (iii) the central value of the gravitational
potential, and (iv) the anisotropy radius, beyond which the velocity
dispersion tensor becomes increasingly radial. As \citet{meylan01}
pointed out, good models are considered as such not only on the basis
of the minimum $\chi^2$ of the surface brightness fit, since the
topology of the $\chi^2$ of the surface has no unique minimum, but
also on the basis of their predictions of the integrated luminosity
and mass-to-light ratio of the clusters. Therefore, \citet{meylan01}
first computed about 150,000 models to explore the parameter space
defined by the IMF exponent, the central gravitational potential, and
the anisotropy radius. They then selected 12 models with the lowest
$\chi^2$ and fulfilling the two requirements above. Since the velocity
dispersion profile for G1 is reduced to one single value, i.e. the
central velocity dispersion, the models are not constrained strongly,
and equally good fits are obtained for rather different sets of
parameters. \citet{meylan01} emphasized that reliable results only
relate to parameters such as the concentration and the total mass, but
probably fail in any more detailed parameters.

\begin{table*}
\caption{G1: Surface brightness, $\mu$, and integrated magnitude,
$m$, as a function of the radius {\bf in the F555 band}}
\begin{tabular}{ccc|ccc}
\hline\hline
   $R$    &  $\mu$  &   $m$   &   $R$    &  $\mu$  &   $m$  \\
 (arcsec) &  (mag)  &  (mag)  & (arcsec) &  (mag)  &  (mag) \\
\hline
\hline
   0.039 & 13.525 $\pm$    0.001 &   19.123 &    0.856 & 16.401 $\pm$    0.005 &   14.583\\
   0.043 & 13.536 $\pm$    0.001 &   19.123 &    0.925 & 16.684 $\pm$    0.004 &   14.536\\
   0.046 & 13.549 $\pm$    0.001 &   19.123 &    0.999 & 16.849 $\pm$    0.005 &   14.484\\
   0.050 & 13.564 $\pm$    0.001 &   19.123 &    1.078 & 16.943 $\pm$    0.005 &   14.433\\
   0.054 & 13.580 $\pm$    0.001 &   18.752 &    1.165 & 16.997 $\pm$    0.005 &   14.383\\
   0.058 & 13.600 $\pm$    0.001 &   18.262 &    1.258 & 17.145 $\pm$    0.008 &   14.335\\
   0.063 & 13.618 $\pm$    0.001 &   18.262 &    1.358 & 17.304 $\pm$    0.011 &   14.292\\
   0.068 & 13.639 $\pm$    0.001 &   18.262 &    1.467 & 17.647 $\pm$    0.005 &   14.253\\
   0.073 & 13.664 $\pm$    0.002 &   18.089 &    1.585 & 17.743 $\pm$    0.006 &   14.214\\
   0.079 & 13.691 $\pm$    0.002 &   17.947 &    1.711 & 17.904 $\pm$    0.007 &   14.178\\
   0.085 & 13.717 $\pm$    0.002 &   17.711 &    1.848 & 17.998 $\pm$    0.006 &   14.142\\
   0.092 & 13.746 $\pm$    0.002 &   17.524 &    1.996 & 18.271 $\pm$    0.007 &   14.108\\
   0.099 & 13.777 $\pm$    0.003 &   17.524 &    2.156 & 18.426 $\pm$    0.007 &   14.075\\
   0.107 & 13.815 $\pm$    0.003 &   17.243 &    2.328 & 18.618 $\pm$    0.006 &   14.043\\
   0.116 & 13.857 $\pm$    0.002 &   17.129 &    2.514 & 18.815 $\pm$    0.008 &   14.014\\
   0.125 & 13.901 $\pm$    0.002 &   16.986 &    2.716 & 18.970 $\pm$    0.007 &   13.986\\
   0.135 & 13.945 $\pm$    0.003 &   16.825 &    2.933 & 19.029 $\pm$    0.007 &   13.960\\
   0.146 & 13.986 $\pm$    0.003 &   16.723 &    3.167 & 19.511 $\pm$    0.006 &   13.936\\
   0.157 & 14.038 $\pm$    0.003 &   16.635 &    3.421 & 19.688 $\pm$    0.009 &   13.915\\
   0.170 & 14.100 $\pm$    0.003 &   16.481 &    3.695 & 19.856 $\pm$    0.007 &   13.894\\
   0.184 & 14.177 $\pm$    0.003 &   16.319 &    3.990 & 19.992 $\pm$    0.013 &   13.873\\
   0.198 & 14.276 $\pm$    0.003 &   16.250 &    4.309 & 20.236 $\pm$    0.012 &   13.855\\
   0.214 & 14.373 $\pm$    0.003 &   16.104 &    4.654 & 20.471 $\pm$    0.011 &   13.838\\
   0.231 & 14.455 $\pm$    0.003 &   15.975 &    5.026 & 20.615 $\pm$    0.010 &   13.821\\
   0.250 & 14.529 $\pm$    0.003 &   15.906 &    5.428 & 20.942 $\pm$    0.014 &   13.804\\
   0.270 & 14.612 $\pm$    0.003 &   15.781 &    5.863 & 21.322 $\pm$    0.011 &   13.791\\
   0.291 & 14.705 $\pm$    0.002 &   15.684 &    6.332 & 21.391 $\pm$    0.016 &   13.778\\
   0.315 & 14.783 $\pm$    0.003 &   15.576 &    6.838 & 21.735 $\pm$    0.018 &   13.766\\
   0.340 & 14.876 $\pm$    0.003 &   15.482 &    7.385 & 21.930 $\pm$    0.022 &   13.755\\
   0.367 & 14.984 $\pm$    0.003 &   15.391 &    7.976 & 22.491 $\pm$    0.015 &   13.746\\
   0.397 & 15.114 $\pm$    0.003 &   15.296 &    8.614 & 22.862 $\pm$    0.012 &   13.736\\
   0.428 & 15.248 $\pm$    0.004 &   15.205 &    9.303 & 22.639 $\pm$    0.028 &   13.727\\
   0.463 & 15.362 $\pm$    0.004 &   15.123 &   10.048 & 22.924 $\pm$    0.020 &   13.718\\
   0.500 & 15.474 $\pm$    0.003 &   15.051 &   10.851 & 23.126 $\pm$    0.028 &   13.708\\
   0.539 & 15.595 $\pm$    0.004 &   14.971 &   11.720 & 23.500 $\pm$    0.022 &   13.700\\
   0.583 & 15.756 $\pm$    0.003 &   14.901 &   12.657 & 23.798 $\pm$    0.018 &   13.693\\
   0.629 & 15.845 $\pm$    0.003 &   14.831 &   13.670 & 23.736 $\pm$    0.021 &   13.685\\
   0.680 & 15.934 $\pm$    0.003 &   14.765 &   14.763 & 23.992 $\pm$    0.029 &   13.678\\
   0.734 & 16.101 $\pm$    0.003 &   14.702 &   15.944 & 24.476 $\pm$    0.040 &   13.674\\
   0.793 & 16.290 $\pm$    0.004 &   14.642 &   17.220 & 24.704 $\pm$    0.049 &   13.670\\
\hline
\end{tabular}
\end{table*}

\subsection{Comparison to previous results}

This cluster has been studied previously by \citet{pritchet84},
\citet{rich96}, \citet{meylan01} and \citet{bk02}. Van den Bergh
(1984) found that the brightest GCs in a number of cluster systems are
also the flattest. To check this conclusion, \citet{pritchet84}
measured the flattening of G1 in the $B$ band, using the CCD camera on
the Canada-France-Hawaii Telescope. Their results showed that G1 is
quite flattened with $\epsilon = 0.22 \pm 0.02$ in the radial range
between $\sim 3$ and $10\arcsec$.  These authors also found that an
empirical King model fitted the surface brightness distribution of G1
very well. Based on {\sl HST}/WFPC2 imaging in F555W with G1 projected
onto the PC (with a pixel size of $0.045\arcsec$), \citet{rich96}
presented photometry of G1, and determined the structural parameters
with the single-mass King models \citep{king66},
$r_c=0.170\pm0.011\arcsec$ ($0.54 \pm 0.04$ pc) and
$r_t=28.21\pm0.44\arcsec$ ($90.0 \pm 1.4$ pc), and $r_h=0.70\arcsec$
and the central surface brightness at $\mu(0)=13.5$ mag
arcsec$^{-2}$. \citet{rich96} found a mean ellipticity $\epsilon\simeq
0.25 \pm 0.02$, which they stated was constant to the core, with no
isophote rotation (P.A. = 122$^\circ$). \citet{rich96} appear not to
have corrected for the instrumental point-spread function (PSF), and
do not state which radial variable they used. Since their ellipticity
and P.A.  measurements are based on elliptical isophote fits, however,
it is likely that the radii they use are in fact the projected
semimajor axis radii as well. Also using {\sl HST}/WFPC2 imaging in
F555W, \citet{meylan01} published aperture photometry of G1, and
determined the structural parameters with multi-mass King models as
defined by \citet{gg79} as follows: $r_c=0.14\arcsec$ (0.52 pc), $r_t
\simeq 54\arcsec$ (200 pc), $r_h=3.7\arcsec$ (14 pc), with a central
surface brightness $\mu(0)=13.47$ mag arcsec$^{-2}$ and a
concentration $c = \log (r_t/r_c) \simeq 2.5$. Although they do not
state the uncertainties in their fits, they use different models to
fit the cluster's surface brightness profile, so that the variation in
the resulting parameters gives us an indication of the associated
uncertainties: $\sigma_{r_c} \simeq 0.01$ pc, $\sigma_{r_t} \simeq 20$
pc, $\sigma_{r_h} \simeq 0.7$ pc, and $\sigma_c \simeq 0.05$. The mean
ellipticity of \citet{meylan01} is $\epsilon\simeq 0.2$. It is evident
that, $r_t$ and $r_h$ of \citet{meylan01} are much larger than those
suggested by \citet{rich96} and in this paper. The reason for this
difference may be that \citet{meylan01} used multi-mass King models,
whereas both \citet{rich96} and the present paper employed single-mass
King models. However, \citet{gg79} have suggested that it is
reasonable for $r_t$ in multi-mass models to differ by a factor of two
from that in single-mass model \citep[also see][]{bk02}. \citet{bk02}
also determined the structural parameters of G1 with single-mass King
models based on the archival {\sl HST}/WFPC2 images in F555W. Their
results showed that $r_c=0.21\arcsec$, $r_t=10.5\arcsec$,
$r_h=0.82\arcsec$, and yielded a central surface brightness
$\mu(0)=13.65$ mag arcsec$^{-2}$. The mean ellipticity was
$\epsilon\simeq 0.20$. \citet{bk02} used the effective radius
[$R_e=(ab)^{1/2}=a(1-\epsilon)^{1/2}$] as the radial variable in the
fits, and they also convolve the fitted model with the instrumental
PSF. As \citet{bk02} pointed out, the resulting scale radii are
systematically larger (by $0.076\pm0.013\arcsec$) and the
concentrations smaller (by $0.09\pm0.02$) than when fitting models
without PSF convolution.

\section{Globular clusters, stripped cores and dwarf spheroidal galaxies}

The distribution of stellar systems in the $M_V$ vs. $\log R_h$ plane
can provide interesting information on the evolutionary history of
these objects \citep[e.g.][]{bergh04,mackey05}. From ground-based
observations of the brightest objects in NGC 5128, the nearest giant
elliptical galaxy, \citet{gomez05} concluded that ``clusters form a
continuum in this diagram''. However, such a conclusion should be
regarded with some caution because (i) the clusters in NGC 5128 were
found to have characteristic half-light radii of $0.3\arcsec$ to
$1\arcsec$, which is only marginally larger than the $0.3\arcsec$ to
$0.6\arcsec$ seeing affecting their observations. Furthermore, (ii)
\citet{gomez05} estimated that roughly 10 per cent of the objects in
their sample might actually be background elliptical galaxies. In view
of these caveats we prefer to restrict our discussion of the
distribution of objects in the $M_V$ vs. $\log R_h$ plane to systems
for which we have access to well-resolved GC-like objects, as long as
their stellar populations are older than a few $\times 10^9$ yr.

Recently, \citet{bergh04} and \citet{mackey05} showed that in a
plot of luminosity versus half-light radius, the overwhelming
majority of GCs, in the Milky Way, the Magellanic Clouds, the
Fornax and Sagittarius dwarf spheroidal galaxies (dSphs), lie
below (or to the right) of the line

\begin{equation}
\log R_{h} ({\rm pc}) = 0.25 M_{V} ({\rm mag}) + 2.95.
\end{equation}

Exceptions to this rule are massive clusters, such as M54 and
$\omega$ Centauri in the Milky Way, and G1 in M31, which are
widely believed \citep{zinnecker88,freeman93,meylan01} to be the
remnant cores of now defunct dwarf galaxies. Because the
well-known giant GC NGC 2419 \citep{bergh04} in the Galaxy and
037-B327 \citep{ma06} in M31 also lie above this line, it has been
speculated that these objects might also be the remnant cores of
dwarf galaxies \citep[but see][for doubts regarding NGC
2419]{degrijs05}. However, more recently a number of large, but
somewhat fainter clusters in M31 and NGC 6822 were found to be
located above Eq. (1) as well \citep{huxor05,huxor06,hwang05}.
This raises the question as to whether there might exist an entire
class of objects which, in the $M_V$ vs. $\log R_h$ plane, are
located between true GCs and dSphs, for which $M_V$ and $R_h$
values were published recently by \citet{irwin05}. Some support
for this speculation is provided by the observations of $M_V$ vs.
$\log R_h$ for the 14 brightest clusters in the peculiar nearby
giant elliptical NGC 5128 \citep{martini04}. These authors
speculate that these brightest NGC 5128 GCs might be nucleated
dwarf galaxies, based on their large masses and the observation by
\citet{harris02} that some show extended envelopes. In addition,
the apparently new class of ultracompact dwarf galaxies (UCDs) in
the Fornax cluster \citep[e.g.][]{mieske02} also occupy a similar
section of parameter space.

\begin{figure}
\resizebox{\hsize}{!}{\rotatebox{-90}{\includegraphics{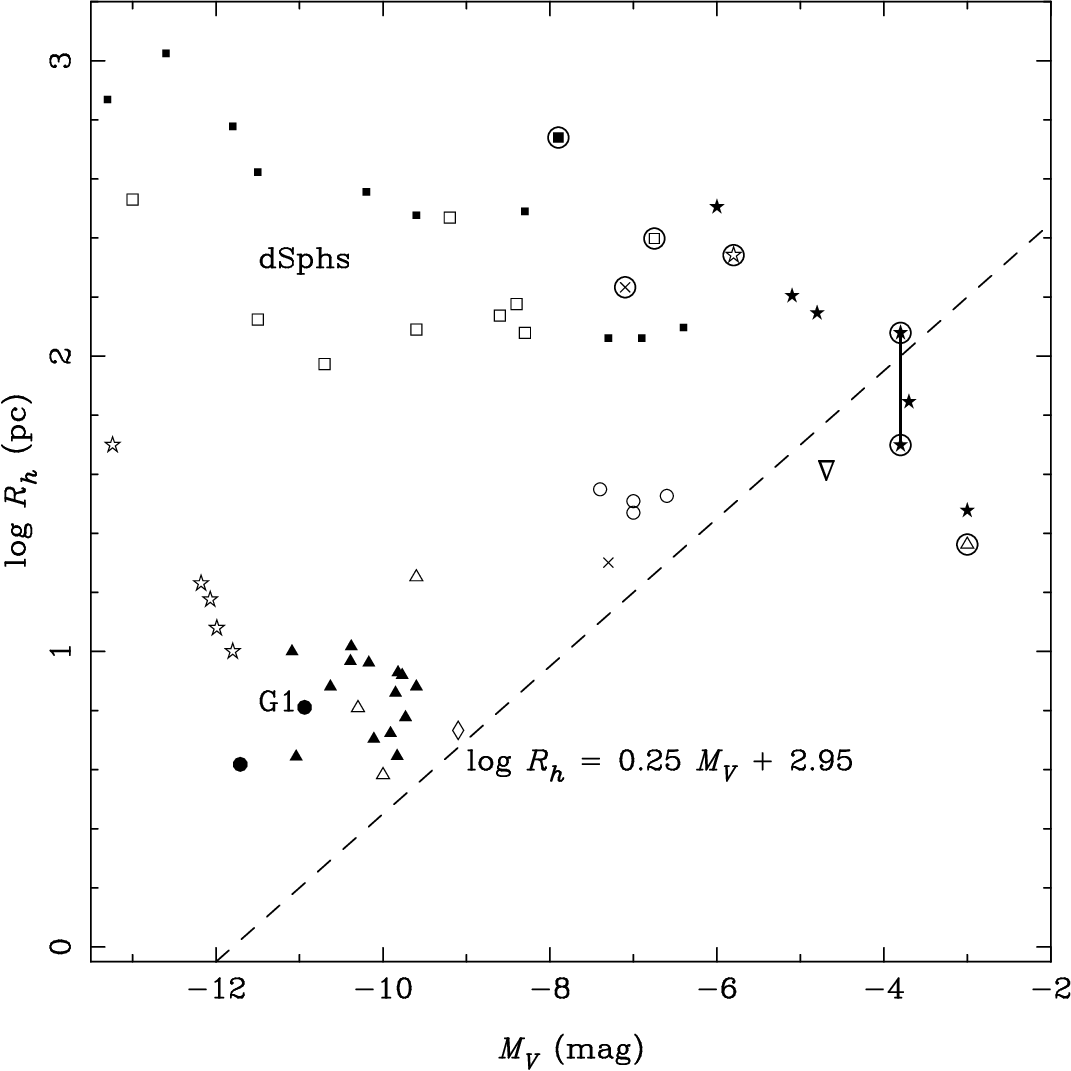}}}
\vspace{0.0cm} \caption{$M_V$ vs. $R_h$ for GCs in M31 (037-B327
and G1: filled circles; B154: open diamond), \citet{huxor05} and
\citet{huxor06}'s new faint large clusters in M31 (open circles),
Galactic dSphs \citep[][open squares]{irwin95}, putative Galactic
stripped dSph cores \citep[][open triangles]{mackey05}, the newly
discovered Milky Way companions (Belokurov et al. 2006: filled
stars; Sakamoto \& Hasegawa 2006: open upside down triangle;
Willman et al. 2005: circled triangle; Willman et al. 2006:
circled open square; Zucker et al. 2006a: circled filled square;
Belokurov et al. 2006: circled open star; Zucker et al. 2006a:
circled filled star; Irwin et al. 2007: circled cross), NGC 6822
GC \citep[][cross]{hwang05}, dSphs associated with the Andromeda
galaxy \citep[][filled squares]{irwin05,harbeck05,martin06}, UCDs
in the Fornax cluster \citep[][stars]{mieske02}, and the brightest
GCs in NGC 5128 \citep[][filled triangles]{martini04}. Also shown
is the line defined by Eq. (1), which gives the upper bound to the
location of normal GCs in the $M_V$ vs. $\log R_h$ plane. The
question as to whether or not the objects shown in this plot form
a continuum is discussed in Sect. 4 of the present paper.}
\label{fig4}
\end{figure}

With the updated value of $R_h$ for G1 from this paper, we
present a plot of $M_V$ vs. $\log R_h$ in Fig. 4, in which $M_V$
was taken from \citet{meylan01}. It is evident that, with the
updated $R_h$, G1 is still seen to lie above and brightward of the
line defined by Eq. (1) \citep{mackey05}. Combined with the
results of \citet{meylan01} that there exists an intrinsic
metallicity dispersion amongst the stars of G1, this strengthens
the conclusion that G1 may be the stripped core of a former dwarf
galaxy \citep[see details from][]{meylan01,mackey05}.
Furthermore, and for completeness, in Fig. \ref{fig4} we have also
included the newly discovered Milky Way companions, based on Sloan
Digital Sky Survey data. These include the objects found by
Belokurov et al. (2006; four probable new dwarf galaxies and one
unusually extended GC, Segue 1), a faint old stellar system at a
distance of $\sim 150$ kpc (Sakamoto \& Hasegawa 2006), which
might either be a new dwarf galaxy or an extended GC, an old,
metal-poor stellar system at a distance of $45\pm 10$ kpc
\citep{willman05a}, which is either an unusual GC or an extreme
dwarf satellite, two new dwarf satellites, one in the
constellation of Ursa Major \citep{willman05b}, and another in the
constellation of Canes Venatici \citep{zucker06a}, a faint new
satellite in the constellation of Bootes at a distance of $\sim
60$ kpc \citep{belokurov06a}, and the faintest known satellite
galaxy in the constellation of Ursa Major \citep{zucker06b}, which
was subsequently confirmed with Subaru imaging, and an unusual
dwarf galaxy in the outskirts if the Milky Way, which lies at a
distance of $\sim 420$ kpc \citep{irwin07}. Finally, the figure
also includes the remote M31 GC B154 (Galleti et al. 2006), for
which S. Galleti kindly provided us with its half-light radius,
$r_h \simeq 1.64$ arcsec.

These results might be taken to suggest (see Fig. \ref{fig4} and Table
3) that dwarf dSphs, stripped cores like $\omega$ Centauri, and normal
GCs either form a continuum in the $M_V$ vs. $\log R_h$ plane, or --
as seems to be the case based on the current best available data --
that the data hint at a possible dichotomy between GCs and stripped
dSph cores on the one hand, and genuine dSphs on the other. A possible
argument supporting latter point of view is that some dSphs, such as
Fornax and Sagittarius, have their own systems of GC companions. Thus,
taken at face value, this result strongly suggests that dSphs and GCs
form systems of different order.  Additional arguments in favour of
this view are that (i) all dSphs appear to contain large amounts of
dark matter, whereas such dark matter seems to be absent from GCs
\citep{pryor89,moore96}.

\begin{table*}
\caption{Data on globular clusters and dwarf galaxies}
\begin{tabular}{llll}
\hline \hline
Object & $M_V$ & $R_h$ & Reference\\
\hline
N2419   &   -9.6   &  17.88 pc   &   Galaxy: \citet{mackey05}\\
N5139  &    -10.3  &   6.44      &                           \\
N6715  &    -10.0  &   3.82      &                           \\
\hline
Carina &    -8.6   &   137 pc    &   \citet{irwin95}         \\
Draco  &    -8.3   &   120       &                           \\
Fornax &    -13.0  &   339       &                           \\
Leo I  &    -11.5  &   133       &                           \\
Leo II &    -9.6   &   123       &                           \\
Sculptor &  -10.7  &    94       &                           \\
Sextans  &  -9.2   &   294       &                           \\
Ursa Minor & -8.4   &   150       &                           \\
\hline
Coma Berenices & -3.7 &  70 pc   &   New satellites of the Milky Way: \citet{belokurov06b}     \\
Canes Venatici II  & -4.8 & 140  &                           \\
Segue 1        & -3.0 &  30      &                           \\
Hercules       & -6.0 &  320     &                           \\
Leo IV         & -5.1 &  160     &                           \\
\hline
SDSS J1257+3419    & -4.8 &  38 pc & A faint old system: \citet{sakamoto06}  \\
\hline
B327   &    -11.71  &   4.15 pc   &    M31: \citet{ma06}\\
\hline
G1     &    -10.94  &   6.5 pc   &    M31: $M_V$ from \citet{meylan01}, and $R_h$ from This paper  \\
\hline
B514    &   -9.1   &   5.41 pc    &   M31: \citet{gall06} \\
\hline
EC1    &    -7.4   &     35.4 pc   &    M31:  \citet{huxor05} and \citet{huxor06}  \\
EC2    &    -7.0   &     29.5      &                           \\
EC3    &    -7.0   &     32.3      &                           \\
EC4    &    -6.6   &     33.7      &                           \\
\hline
SC1    &    -7.3   &     20 pc   &  NGC 6822: \citet{hwang05}\\
\hline
And I  &    -11.8  &   0.60 kpc  &  M31 companions: \citet{irwin05}\\
And II &    -12.6  &   1.06      &                           \\
And III&    -10.2  &   0.36      &                           \\
And V  &     -9.6  &   0.30      &                           \\
And VI &    -11.5  &   0.42      &                           \\
And VII&    -13.3  &   0.74      &                           \\
\hline
And IX &     -8.3  &   0.31 kpc  & M31 companion: \citet{harbeck05}\\
\hline
And XI &     -7.3  &   115 pc  & M31 companions: \citet{martin06}\\
And XII &    -6.4  &   125 pc  &                                 \\
And XIII &   -6.9  &   115 pc  &                                 \\
\hline
SDSSJ$1049+5103$ & -3.0 & 23 pc & Milky Way companion: \citet{willman05a}\\
\hline
Ursa Major & -6.75 & 250 pc & New dwarf galaxy of the Milky Way: \citet{willman05b}\\
\hline
Canes Venatici & -7.9 & 550 pc & New dwarf satellite of the Milky Way: \citet{zucker06a}\\
\hline
Bo{\"o}tes & -5.8 & 220 pc & New faint satellite of the Milky Way: \citet{belokurov06a}\\
\hline
Ursa Major II & -3.8 & 50 pc or 120 pc & New curious satellite of the Milky Way: \citet{zucker06b}\\
\hline
Leo T         & -7.1 & 171 pc &  An unusual dwarf galaxy in
the outskirts of the Milky Way: \citet{irwin07}\\
\hline
\end{tabular}
\end{table*}

However, a weakness of this argument is that dark matter that once
may have surrounded GCs might have been stripped from them by
tidal interactions \citep[e.g.,][]{saito05}. Furthermore, (ii)
individual stars in GCs (with the notable exceptions of $\omega$
Centauri, {\bf M54} and possibly M22 in the Galaxy and G1 in M31)
all have similar metallicities, whereas individual stars in dSphs
exhibit a wide range in [Fe/H] values. Finally, \citet{pritzl05}
noted that the [$\alpha/\rm Fe$] and light r-process element
ratios in most GCs mimic those in stars of similar metallicity in
the Galactic field, and differ from those in dwarf galaxies. Thus,
the apparent dichotomy in the $M_V$ vs. $\log R_h$ plane shown in
Fig. \ref{fig4} might well be a distinction in dark matter content
and approximate coevality of the stellar content of these objects.
This view is supported by the observation that while the vast
majority of Galactic and Magellanic Cloud GCs are very nearly
coeval, there are indeed clear metallicity spreads in the objects
lying above the dividing line.

It has also been claimed \citep{carrara06} that the very old and
metal-rich open cluster NGC 6791 might be the remnant core of a
dSph galaxy as well. However, arguments against this hypothesis
include, (i) with $\rm [M/H] = 0.39\pm0.05$, NGC 6791 would be
much more metal-rich than any other nearby putative stripped dwarf
galaxy core, (ii) the small number of cluster stars for which
metallicities have thus far been determined do not exhibit a
significant metallicity spread. In this respect, this object would
therefore differ from $\omega$ Centauri and a number of dSph
companions to the Galaxy. Finally, (iii) it has been suggested by
van den Bergh (2000, pp. 54-55, and references therein) that NGC
6791 was originally a cluster in the metal-rich Galactic bulge
that was ejected into the disc by tidal interactions with the
massive bar believed to be located within the Galactic bulge.
Further support for this suggestion has recently been provided by
means of accurate, high spatial resolution proper motion
measurements based on multi-epoch {\sl HST}/ACS imaging (Bedin et
al. 2006). Some light might be shed on this question if the values
of $M_V$ and $R_h$ were available for NGC 6791. This might allow
one to see if NGC 6791 falls above, or below the line in the $M_V$
vs. $\log R_h$ plane defined by Eq. (1).

In summary it is concluded that the sample of objects that can be
reliably placed in the $M_V$ vs $\log R_h$ diagram is still too small
to decide whether this plane is normally occupied by a continuum of
objects, or if unusual conditions such as ``tidal thrashing'' are
required to fill some regions of this plane. In this connection we
note that the four ultracompact dwarfs (UCDs) in the Fornax cluster
\citep{propris05} for which $M_V$ and $R_h$ values are available (all
of which are located within $30\arcmin=150$ kpc of NGC 1399), lie in a
compact grouping near $\langle M_V \rangle~=~-11.75$, $\langle \log
R_h \rangle~=~1.25$. In Fig. \ref{fig4}, this places these objects
near the centre of the apparent ``zone of avoidance'' between the
Local Group dSphs and the putative stripped cores of dwarf galaxies,
which extends up to $\sim0.7$~dex above the line defined by Eq. (1).
Similarly, \citet{hasegan05} have found three objects in the core of
the Virgo cluster near M87 that lie between the regions in the $M_V$
vs. $\log R_h$ diagram that are usually occupied by normal dSphs and
GC-like objects that are thought to be the stripped cores of such
dSphs.

Finally, as has been pointed out by \citet{irwin05}, it is also
puzzling that there appears to be a systematic difference between the
locations of Galactic and Andromeda dSphs in the $M_V$ vs. $\log R_h$
plane.

\section{Summary}

In this paper, we redetermined the structural parameters of Mayall II
= G1 based on an F555W image obtained with the Advanced Camera for
Surveys on the {\sl HST}, by performing a fit to the surface
brightness distribution of a single-mass isotropic King model. This
allowed us to probe to smaller radii than ever before, thanks to the
significantly higher spatial resolution offered by our instrumental
set-up compared to that used by previous authors. We derive a core
radius, $r_c=0.21\pm0.01\arcsec~(=0.78\pm0.04~\rm{pc})$, a tidal
radius, $r_t=21.78\pm1.06\arcsec~(=80.7\pm3.9~\rm{pc})$, and a
concentration index $c=\log (r_t/r_c)=2.01\pm0.02$. The central
surface brightness is 13.510 mag arcsec$^{-2}$. We calculate the
half-light radius, at $r_h=1.73\pm0.07\arcsec(=6.5\pm0.3~\rm{pc})$.
The results show that, within 10 core radii, a \citet{king66}
model fits the surface brightness distribution well, although a
single-mass King model cannot fit the observed profile at the outer
regions well. The reason for this may be that for G1, which has been
considered as the possible remnant core of a former dwarf elliptical
galaxy, it is impossible to model the complicated stellar and
dynamical properties based on simple theories for GCs, such as King
models. This applies in particular to the outer regions, where there
exist strong tidal force due to the host galaxy. We also discussed
the variation of ellipticity and position angle, and of surface
brightness for the core of the object, in relation to previous
measurements. We find that G1 falls in the same region of the $M_V$
vs. $\log R_h$ plane as $\omega$ Centauri, M54 and NGC 2419 in the
Galaxy. All three of these objects have been claimed to be the
stripped cores of now defunct dwarf galaxies. We discussed in detail
whether GCs, nucleated dSph cores and normal dwarf galaxies form a
continuous distribution in the $M_V$ vs. $\log R_h$ plane, or if GCs
and dSphs constitute distinct classes of objects; we have presented
arguments in favour of this latter option.

\section*{Acknowledgments}
This paper is based on observations made with the NASA/ESA {\sl
Hubble Space Telescope}, obtained at the Space Telescope Science
Institute, which is operated by AURA, Inc., under NASA contract
NAS 5-26555. These observations are associated with proposal 9767.
RdG acknowledges an International Outgoing Short Visit grant to
the National Astronomical Observatories of China in Beijing from
the Royal Society. We are indebted to the referee for his/her
thoughtful comments and insightful suggestions that improved this
paper greatly. We are also indebted to Narae Hwang for providing
us with his data on three GCs in the outer regions of NGC 6822 in
advance of publication. Thanks are also due to Eva Grebel and Alan
McConnachie for their help. We thank Silvia Galleti and Michele
Bellazzini for providing us with the half-light radius
measurements for B154 prior to publication. This research has made
use of the Astrophysical Integrated Research Environment (AIRE),
which is operated by the Center for Astrophysics, Tsinghua
University. This work was supported by the Chinese National
Natural Science Foundation grands No. 10473012, 10573020,
10633020, 10673012, and 10603006.

\end{document}